\documentclass[aps,prb,twocolumn,longbibliography,groupedaddress,showpacs,floatfix,superscriptaddress]{revtex4-1}
\usepackage[colorlinks=true,citecolor=blue,linkcolor=blue,breaklinks=true]{hyperref}
\usepackage{graphicx}  
\usepackage{dcolumn}   
\usepackage[english]{babel}
\usepackage{bm}        
\usepackage{amsmath, amsthm, amssymb}
\usepackage{bbold} 
\usepackage{mathrsfs}
\usepackage{array}
\usepackage[usenames]{color}
\usepackage{soul} 
\usepackage{ulem} 
\usepackage{float}

\emergencystretch=\maxdimen
\begin{document}
\title
{Effects of lattice geometry on thermopower properties of the repulsive Hubbard model}
\author{Willdauany C. de Freitas Silva}
\affiliation{Instituto de F\'isica, Universidade Federal do Rio de Janeiro, Rio de Janeiro, RJ 21941-972, Brazil}
\author{Maykon V. Araujo}
\affiliation{Departamento de F\'isica, Universidade Federal do Piau\'i, 64049-550 Teresina PI, Brazil}
\author{Sayantan Roy}
\affiliation{Department of Physics, The Ohio State University, Columbus OH 43210, USA}
\author{Abhisek Samanta}
\affiliation{Department of Physics, The Ohio State University, Columbus OH 43210, USA}
\author{Natanael de C. Costa}
\affiliation{Instituto de F\'isica, Universidade Federal do Rio de Janeiro, Rio de Janeiro, RJ 21941-972, Brazil}
\author{Nandini Trivedi}
\affiliation{Department of Physics, The Ohio State University, Columbus OH 43210, USA}
\author{Thereza Paiva}
\affiliation{Instituto de F\'isica, Universidade Federal do Rio de Janeiro, Rio de Janeiro, RJ 21941-972, Brazil}


\begin{abstract}
We obtain the Seebeck coefficient or thermopower $S$, which determines the conversion efficiency from thermal to electrical energy, for the two-dimensional Hubbard model on different geometries (square, triangular, and honeycomb lattices) for different electronic densities and interaction strengths. Using Determinantal Quantum Monte Carlo (DQMC) we find the following key results: (a) the bi-partiteness of the lattice affects the doping dependence of $S$; (b) strong electronic correlations can greatly enhance $S$ and produce non-trivial sign changes as a function of doping especially in the vicinity of the Mott insulating phase; (c) $S(T)$ near half filling can show non-monotonic behavior as a function of temperature.  
We emphasize the role of strong interaction effects in engineering better devices for energy storage and applications, as captured by our calculations of the power factor $PF=S^2 \sigma$ where $\sigma$ is the dc conductivity. 

\end{abstract}


\maketitle

\section{Introduction}
Over the past decades, a great deal of interest has been given to increasing the efficiency of electrical devices. 
As a possible route to this end, the thermoelectric materials may play a crucial role, once they exhibit induced voltage in presence of a temperature gradient, whose magnitude is quantified by the Seebeck coefficient, the thermoelectric Power Factor, and, eventually, the Figure of Merit\,\cite{H_bert_2015}.
However, there are many technical issues that make the development of efficient thermoelectric materials a challenge, e.g., the toxicity of the compounds, or their thermal instability.
There are few ways to overcome these problems: (i) optimizing the already known compounds through band-structure engineering and nanostructuration techniques\,\cite{Guangqian_2015,Joseph_2008,Hicks_1993}, or (ii) seeking new classes of compounds which exhibit unconventional properties 
usually related to strong electron-electron interactions\,\cite{Wiss_2010}.
In view of the increasing number of novel correlated compounds, controlling and manipulating 
geometry and correlations to enhance the thermopower properties is an open issue.


Despite intense experimental efforts, further theoretical investigations are required, in particular to understand interacting electronic compounds, such as Na$_{x}$CoO$_{2}$ or FeSb$_{2}$, which exhibit unusual large thermopower response\,\cite{Wiss_2010,Wiss_2011,Tomczak_2018}.
In the former, the combination of quasi two-dimensional character with band topology and strong electronic correlations make this material an interesting playground to examine thermoelectricity. Once the charge carriers are confined to the hexagonal layers of Co atoms, a disordered distribution of Na ions above and below it can induce a charge imbalance. Using local-density approximation and dynamical mean-field theory, Held et al\,\cite{Wiss_2010} showed that  disorder combined with the pudding-mold band structure and strong correlations enforce the electron-hole imbalance and enhances the thermopower by $200 \% $ with respect to the non-interacting case.

The thermopower of superconducting cuprates has been experimentally studied\,\cite{high-tc-1,high-tc-2,high-tc-3,high-tc-4}, with different compounds displaying very similar behavior with a sign change of the Seebeck coefficient near the maximum critical temperature\,\cite{high-tc-1}. This nearly universal behavior has been the subject of theoretical interest, being ascribed to a possible underlying critical point\,\cite{high-tc-teo-3}, to the presence of a van Hove singularity\,\cite{high-tc-teo-1}, and has been recently observed for the Hubbard Model with next-nearest neighbor hopping\,\cite{Deveraux}.


In view of these stimulating results, we explore how  electron-electron interactions and geometry  affect the Seebeck coefficient, and the thermoelectric Power Factor.
To this end, we use unbiased Quantum Monte Carlo simulations to study the single band repulsive Hubbard model. We analyze the thermoelectric and electrical transport properties in the long wavelength DC limit in two-dimensions, in the square, triangular and honeycomb lattices. Our study finds a strong influence of particle hole symmetry of the many body spectrum and DOS on the behavior of the Seebeck coefficient with respect to doping. We show non trivial sign changes of the Seebeck coefficient that signal a deviation of the Fermi surface from the Luttinger count and a subsequent anomalous change of the type of carriers below and above half filling. The sign change is also accompanied by a significant increase of the Seebeck coefficient near half filling with respect to the non-interacting and weakly interacting case. We show that despite using a simplified thermodynamic formula for the Seebeck coefficient that is independent of dynamical quantities, we are able to capture the effects of strong correlation.

The paper is organized as follows: In Section \ref{model_meth} we discuss the Hubbard model, an introduction to the Seebeck coefficient, Kelvin formula and the auxiliary field QMC method used to solve it. In Section \ref{Entropy} we discuss how to calculate the entropy and present our results for this quantity. Section \ref{DOS_conductivity} shows the Local Density of States and conductivity results. In Section \ref{Sec_Seebeck}  we discuss the Seebeck coefficient and  in Section \ref{Sec_Power_factor} the Power Factor. Finally, in section \ref{conclu} we summarize our findings.

\section{Model and methods}
\label{model_meth}
The repulsive Fermi Hubbard model describes electrons on a lattice with an onsite repulsive interaction, with the Hamiltonian
\begin{align}\label{Eq:Hamil}
\nonumber \mathcal{H} = & -t\sum_{\substack{\langle \textbf{i},\textbf{j} \rangle},\sigma} \big( c_{\textbf{i}, \sigma}^{\dagger}c_{\textbf{j}, \sigma}+ {\rm H.c.} \big) - \mu \sum_{\substack{\textbf{i}}, \sigma} n_{\textbf{i},\sigma}
\\  & + U   \sum_{\substack{\textbf{i}}} (n_{\textbf{i},\uparrow}-1/2)(n_{\textbf{i},\downarrow}-1/2),
\end{align}
where the sums run over sites of a given two-dimensional lattice, with $\langle \mathbf{i}, \mathbf{j} \rangle$ denoting nearest-neighbor sites.
Here, we use the second quantization formalism, with $c^{\dagger}_{\mathbf{i}, \sigma}$ ($c^{\phantom{\dagger}}_{\mathbf{i}, \sigma}$) being creation (annihilation) operators of electrons on a given site $\mathbf{i}$, and spin $\sigma$, while $n_{\mathbf{i},\sigma} \equiv c^{\dagger}_{\mathbf{i}, \sigma} c_{\mathbf{i}, \sigma}$ are number operators.
The first two terms on the right-hand side of Eq.\,\eqref{Eq:Hamil} correspond to the hopping of electrons, and the chemical potential $\mu$, respectively, with the latter determining the filling of the lattice.
The third term describes the local repulsive interaction between electrons, with interaction strength \textit{U}; the factor of $1/2$  is introduced to ensure invariance of the hamiltonian under particle-hole transformations on bipartite lattices. This implies that for the bipartite lattices we consider here (square and honeycomb lattices), $\mu=0$ sets half-filling for all temperatures.

Our central quantities of interest are the transport coefficients, and their behavior with respect to doping and strength of interactions. The transport coefficients are defined through the following relations,
\begin{align}
    \vec{j} &= \tensor{L}^{11}\vec{E}+\tensor{L}^{12}(-\vec{\nabla} T) \nonumber \\ 
    \vec{j}^q &= \tensor{L}^{21}\vec{E}+\tensor{L}^{22}(-\vec{\nabla}T)
    \label{Eq:conductivity}
\end{align}
where $\vec{j}$ and $\vec{j}^q$ are the electrical and thermal currents, and the $\tensor{L}$s are rank 2 tensors defining conductivities of the system. The tensors $\tensor{L}^{11}$ and $\tensor{L}^{22}$ are the electrical and thermal conductivities, and $\tensor{L}^{12}$($\tensor{L}^{21}$) are the thermoelectrical(electrothermal) conductivities. The thermopower or Seebeck coefficient is defined as
\begin{align}
    S &= \frac{\big(\tensor{L}^{12}\big)_{xx}}{\big(\tensor{L}^{11}\big)_{xx}}= \frac{1}{T}\frac{\big(\tensor{L}^{21}\big)_{xx}}{\big(\tensor{L}^{11}\big)_{xx}},
    \label{Eq:Seebeck}
\end{align}
where the second equality is due to Onsager's reciprocity relations\cite{onsager1931reciprocal}.
Using linear response theory with respect to electrical field and temperature, the Seebeck coefficient in the Kubo formalism can also be written as

\begin{align}
    S(q_x,\omega) =  \frac{1}{T}\frac{\chi_{\hat{\rho}(q_x),\hat{K}(-q_x)}(\omega)}{\chi_{\hat{\rho}(q_x)\hat{\rho}(-q_x)}(\omega)} 
    \label{Eq:Seebeck_Kubo}
\end{align}
where 
\begin{align}
    \chi_{\hat{\rho}(q)\hat{K}(-q)}(\omega) &\!=\! \lim_{\eta \rightarrow 0}\sum_{n,m}(f_n\!-\!f_m)\frac{\langle n|\hat{\rho}(q)|m\rangle \langle m|\hat{K}(-q)|n\rangle}{\omega+i\eta+\epsilon_n-\epsilon_m} \nonumber \\
        \chi_{\hat{\rho}(q)\hat{\rho}(-q)}(\omega) &= \lim_{\eta \rightarrow 0}\sum_{n,m}(f_n\!-\!f_m)\frac{\langle n|\hat{\rho}(q)|m\rangle \langle m|\hat{\rho}(-q)|n\rangle}{\omega+i\eta+\epsilon_n-\epsilon_m} \nonumber \\
        \label{Eq:Susceptibility}
\end{align}
define the electrothermal and electrical conductivities, respectively. Evaluation of a Kubo-like formula (Eq \eqref{Eq:Seebeck_Kubo}) is not easy for interacting systems in the thermodynamic limit, although it is the most insightful. Alternate formulas like the Mott formula\cite{jonson1980mott}, Heikes-Mott\cite{heikes1961}, high frequency Seebeck\cite{shastry2006sum,shastry2008electrothermal,zdroj2007} and Kelvin formula\cite{peterson2010kelvin,shastry2008electrothermal} exist but are limited by their applicability to specific scenarios (weakly correlated metal at low temperatures for the Mott formula, high-temperature limit for the Heikes-Mott formula, and measurement of transport at high frequencies compared to characteristic energy scale for high-frequency Seebeck formula). The Kelvin formula was proposed by Lord Kelvin to provide reciprocity between Seebeck and Peltier coefficients, 
and is calculated in the slow limit ($q_x \rightarrow 0,\omega \rightarrow 0$); it can be derived by taking the slow limit of Eq (\ref{Eq:Seebeck_Kubo}) \cite{peterson2010kelvin}. 
The Kelvin formula for the Seebeck coefficient is
\begin{align}
\label{eq:seebeck}
\nonumber S_{\rm Kelvin} &= \left. - \frac{1}{e} {\frac{\partial \mu }{\partial T}}\right|_{V,n} = \left.  \frac{1}{e} {\frac{\partial s}{\partial n}}\right|_{T,V}~,
\end{align}
where the second equality follows from Maxwell's relations. Although expressed in terms of thermodynamic quantities, which are sufficient to capture the effects from the many-body density of states, it misses kinematic factors like contributions from velocities at the Fermi surface and relaxation times. Nonetheless, the effects of strong correlations on the low-frequency transport behavior are taken into account, as it retains the $\omega<U$ approximation, which the high frequency formula, $S^{*}$ misses. The justification and benchmark of using the Kelvin formula for strongly interacting systems like the Hubbard model and fractionalized systems like $\nu = 5/2$ FQHE states have already been established\cite{arsenault2013entropy,peterson2010kelvin}.

We investigate the thermodynamics and transport properties of Eq.\,\eqref{Eq:Hamil} on three different lattices: square, triangular and honeycomb. In particular, we examine the behavior of the
entropy, conductivity, Local Density of States (LDOS), Seebeck
coefficient, and Power Factor as functions of lattice filling, for
different values of interaction strengths. To this end, we perform unbiased determinant quantum Monte Carlo (DQMC) simulations~\cite{Blankenbecler81,Hirsch83,Hirsch85,White89},  a state-of-the-art numerical method that maps a many-particle interacting fermionic system into a single-particle (quadratic form) one, with the aid of bosonic auxiliary fields.
More details about the methodology may be found in, e.g., Refs.\,\onlinecite{gubernatis16,becca17,Santos03}, and references therein.
Our DQMC simulations are performed for finite-sized systems (with 100, 144 and 162 sites for the square, triangular, and honeycomb lattices, respectively), and for interaction strengths $U/t = 0$, 2, 4, 6, 8, and 10; i.e., from non-interacting to strong coupling regimes.
Throughout this work, unless otherwise indicated, we consider $T/t=0.5$, which corresponds to an energy scale low enough to observe the crossover towards an insulating phase at half-filling~\cite{Kim20,Simkovic20,Lenihan21}. Hereafter, we define the lattice constant as unity, and the hopping integral $t$ as the energy scale.

\section{Entropy}
\label{Entropy}
\begin{figure}[t]
\includegraphics[scale=0.40]{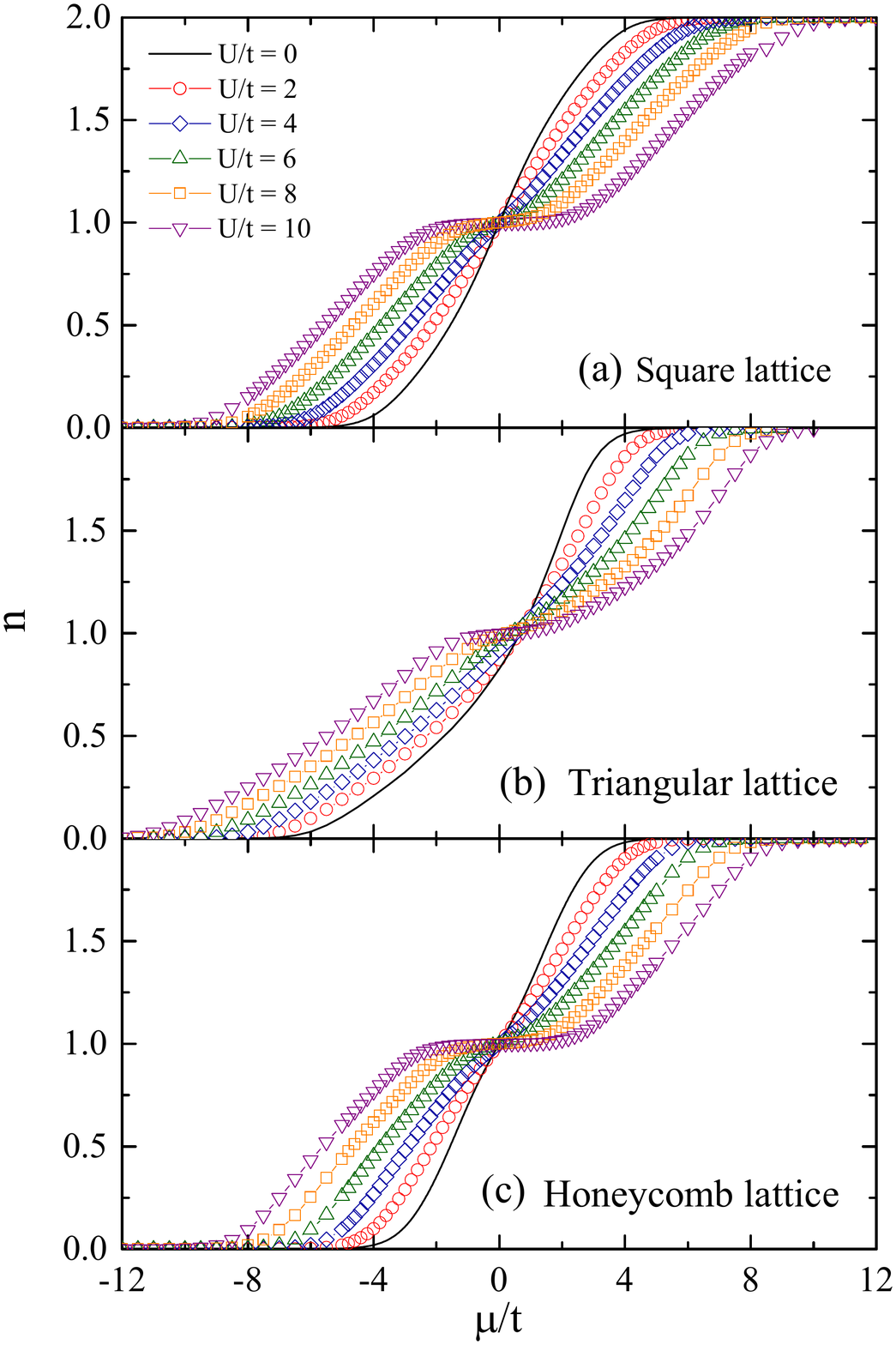}
\caption{(Color online) Density $n$ as a function of chemical potential $\mu$ for the (a) square, (b) triangular, and (c) honeycomb lattices, for fixed $T/t=0.5$, and different interaction strengths $U/t$. Here, and in all subsequent figures, when not shown, error bars are smaller than symbol size. A Mott plateau is formed above a critical interaction strength which depends on lattice geometry. Note the width of this plateau is a measure of the charge gap or ``Mottness".}
\label{fig:density}
\end{figure}

\begin{figure}[t]
\includegraphics[scale=0.40]{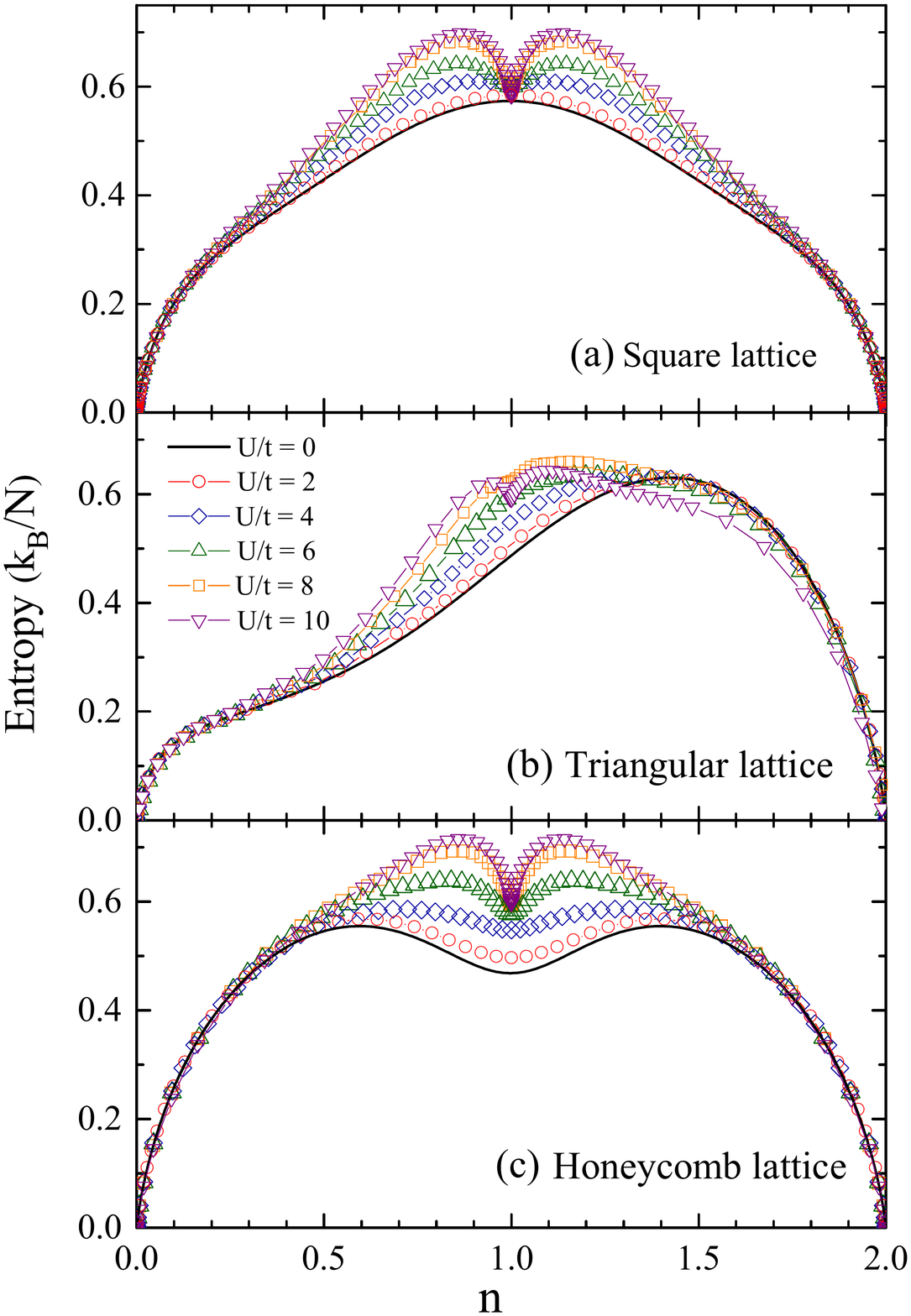}
\caption{(Color online) Entropy $s$ as a function of the electronic density $n$ for the (a) square, (b) triangular, and (c) honeycomb lattices, for fixed $T/t=0.5$, and different interaction strengths $U/t$. Note that the entropy is very different among the three lattices in the noninteracting limit. In presence of sufficiently large interactions, a local
minimum appears at the half-filling for all three lattices; However, it becomes qualitatively similar for the square and honeycomb lattices, while it is different for the non-bipartite triangular lattice.}
\label{fig:entropy}
\end{figure}

We start our analysis with the electronic density as a function of chemical potential shown in Figure \ref{fig:density} for $T/t=0.5$ and the different lattices studied: (a) square, (b) triangular, and (c) honeycomb. Particle-hole symmetry is evident for the square and honeycomb lattices, where $n(\mu)=2-n(-\mu)$. In all geometries, correlations lead to the formation of a Mott plateau (insulating phase) around half-filling, although the critical interaction strength for the onset of ``Mottness" 
is strongly influenced by the lattice geometry. 

As the Seebeck coefficient is obtained from the entropy, we now turn to discuss its behavior for these geometries.
The entropy can be obtained from the electronic density, $n(\mu)$,
by integrating it over the chemical potential $\mu $,
\begin{equation}
    s(\mu, T) = \int_{-\infty}^\mu d\mu \left.  \frac{\partial n}{\partial T}\right|_\mu 
\label{eqn:entropy}
\end{equation}
Figure \ref{fig:entropy}\,(a) displays $s(\mu,T)$ as a function of $n$ for the square lattice. One may notice that, at $T/t=0.5$, the results for $U/t = 2$ (red circles symbols) and $U/t = 4$ (blue diamond symbols) exhibit very similar behavior to the non-interacting one (solid black curve), once  the temperature is high enough to destroy the correlation effects for such small interaction strengths. However as
 $U/t$ increases, e.g.~$U/t=6$, 8, and 10 the entropy presents a local minimum at half-filling, being drastically reduced as $T/t \to 0$, due to the Mott gap formation in the ground state\,\cite{Bonca03,Khatami11,Simkovic20,Mikelsons09,Paiva01}. The increased entropy for the metallic state in the vicinity of half-filling in the presence of interactions has been observed for the cubic lattice and is relevant for cold  fermionic atoms trapped in optical lattices, where the metallic region of the atomic cloud is used to absorb entropy and allow a central Mott region at a higher entropy per particle \cite{Paiva11}. 

Unlike what is observed for the square lattice, the entropy behavior on the triangular geometry only exhibits such a local minimum at $U/t=10$, the largest interaction strength considered, as presented in Fig.\,\ref{fig:entropy}\,(b).
This suggests an absence of a Mott gap for weak and intermediate interaction strengths. In fact, at $U/t\gtrsim 8$, a small dip starts to form around $n=1$, in line with the expectation for a metal-to-insulator transition occurring for $U/t \approx 7-8$ \cite{Yoshioka,Shirikawa,Schauss}. The lack of particle-hole symmetry in the triangular lattice  shown in Fig.\,\ref{fig:density}\,(c) is clearly also present in the entropy.

The entropy for the honeycomb lattice is more subtle.
Similar to the square and triangular lattices, a local minimum appears at half-filling in presence of strong interaction; however, different from what is seen in the two previous cases, here $s(n, T)$ displays a suppression at half-filling even for the non-interacting case.
This behavior is understood by recalling that the honeycomb lattice has a vanishing  DOS  at half-filling, with van Hove singularities below and above it.
That is, the results of Fig.\,\ref{fig:entropy} confirm our expectations that the entropy provides hints about the DOS, irrespective of the interaction strengths.
These odd features of the honeycomb lattice lead to strong changes for the Seebeck coefficient at $U=0$, as we shall see in Section \ref{Sec_Seebeck}, and may obscure some of the the electronic correlation effects.

\section{Local Density of states and conductivity}
\label{DOS_conductivity}

As discussed before, Fig.\,\ref{fig:density} signals the occurrence of a Mott insulating state due to the presence of a plateau in the density as a function of chemical potential, driven by correlations. The subtleties in the entropy require an analysis of  the suppression of spectral weight for different densities. In order to avoid numerical analytical continuations,  we examine the LDOS only at the Fermi level ($\omega=0$), which is obtained through\,\cite{Trivedi95}
\begin{equation}
\label{eq:densos}
N(\omega = 0) \approx \frac{\beta}{\pi} G(|\mathbf{i}-\mathbf{j}| = 0,\; \tau = \beta/2),
\end{equation} 
where $G(\mathbf{r}, \tau)$ is the real space and imaginary time Green's function calculated at $\mathbf{r} \equiv \mathbf{i}-\mathbf{j} = 0$, and imaginary time $\tau = \beta/2$ (hereafter, $\beta \equiv 1/(k_BT)$ is the inverse of temperature, and $k_B$ the Boltzmann constant).
\begin{figure}[t]
\includegraphics[scale=0.40]{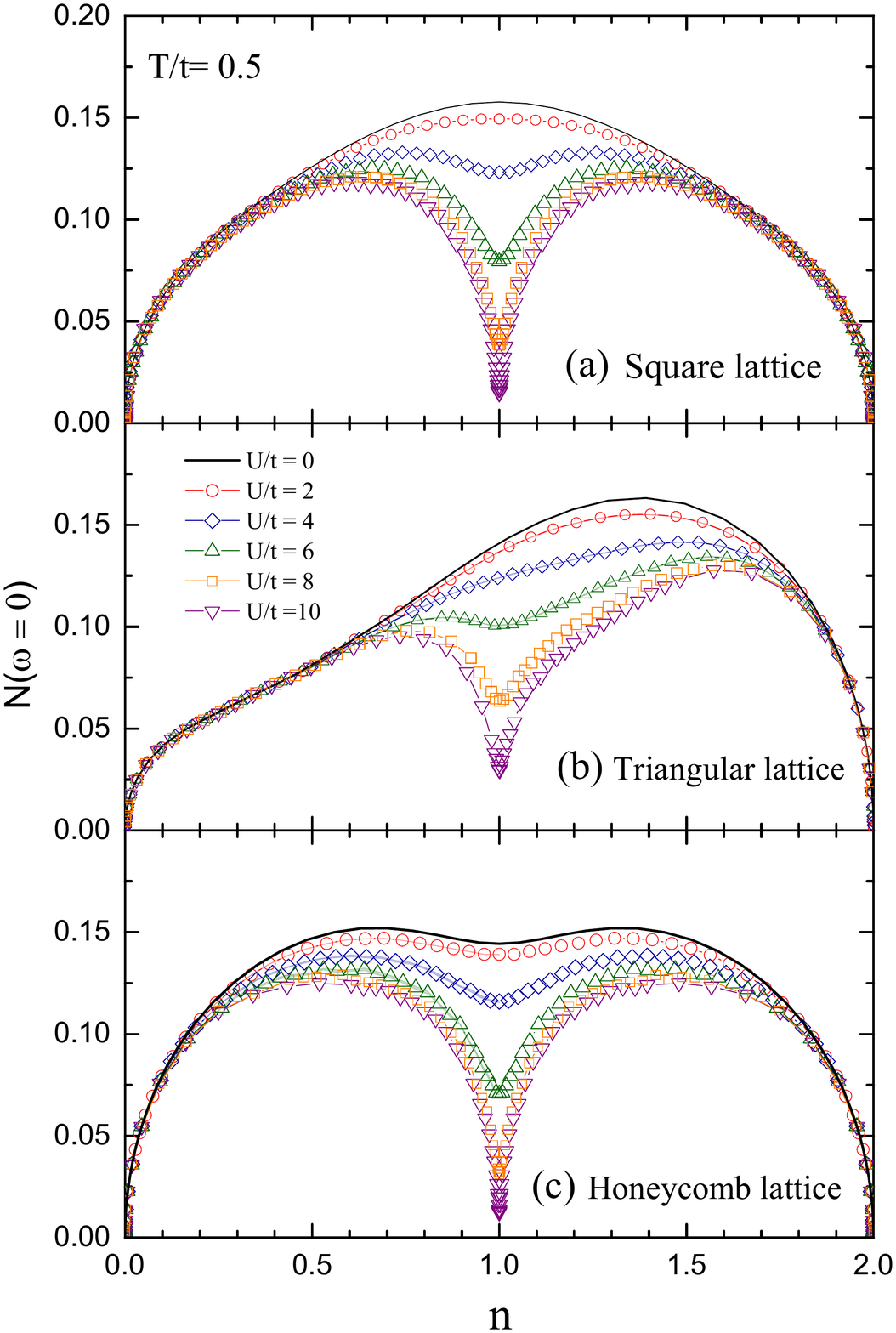}
\caption{(Color online) Local density of states (LDOS) at the Fermi level at $T/t=0.5$ as a function of density for the (a) square, (b) triangular, and (c) honeycomb lattices. Note the similarity of the doping dependence of LDOS to that of the entropy (Fig.~\ref{fig:entropy}) in the non-interacting limit. When correlations increase, LDOS develops a dip near half-filling, leading to maxima at intermediate densities which depend on the lattice geometry. The LDOS also becomes qualitatively similar to the square and honeycomb lattices in the strongly interacting limit, but is quite different from the triangular lattice which, in contrast, is not bipartite and has an asymmetric particle-hole many-body spectrum.}
\label{fig:DOS}
\end{figure}

Fig.~\ref{fig:DOS} shows $N(\omega=0)$ for the (a) square, (b) triangular, and (c) honeycomb lattices. 
 A common feature for all geometries is that the non-interacting case is an upper bound for the LDOS, with data close to the empty and completely filled systems showing a negligible  dependence on the interaction strength. 
For the temperature shown, $T/t=0.5$, we can see that for the square lattice and $U/t=2$ the dip at half-filling has not developed yet. As correlations increase, the dip starts to form, leading to maxima at $n \approx 0.6$ and  $n \approx 1.4$. The non-interacting ground state for the honeycomb lattice is known to be a semi-metal, with a vanishing LDOS at half-filling at $T/t \rightarrow 0$. Figure \ref{fig:DOS}(c) shows a dip in the DOS at $T/t=0.5$ already for $U/t=0$, in line with entropy data. Similar to what is seen for the square lattice, maxima develop as correlations increase, but the positions are moved to $n = 0.5$ and $n=1.5$. The triangular lattice LDOS shows a distinct behavior, as can be seen in figure \ref{fig:DOS}(b), with a broad peak for $U/t=0$ located at $n \approx 1.6$. The effects of correlations are only relevant for $n \gtrsim 0.6$. For the larger values of $U/t$ considered, a small broad peak is present at $n \simeq 0.75$ and a higher one at $n \simeq 1.6$, with a dip at half-filling signaling the Mott state for large $U$.

To further investigate the transport properties, we now turn to the longitudinal dc conductivity,
\begin{equation}\label{eq:sigma_dc}
\sigma_{dc} = \frac{\beta^2}{\pi} \Lambda_{xx}(\mathbf{q=0}, \tau = \beta/2),
\end{equation}
in which
\begin{equation}
\Lambda_{xx}(\mathbf{q}, \tau ) = 
\langle j_{x}(\mathbf{q}, \tau) j_{x}(-\mathbf{q}, 0)  \rangle
\end{equation}
is the current-current correlation function with $ j_{x}(\mathbf{q}, \tau) $ being the Fourier transform of the unequal-time current operator
\begin{equation}
j_x(\mathbf{i},\tau)=\mathrm{e}^{\tau\mathcal{H}}
  \left[
        it\sum_\sigma
            \left(c_{\mathbf{i}+\mathbf{x},\sigma}^\dagger 
                  c_{\mathbf{i},\sigma}^{\phantom{\dagger}}
                  - 
                  c_{\mathbf{i},\sigma}^\dagger  
                  c_{\mathbf{i}+\mathbf{x},\sigma}^{\phantom{\dagger}}
            \right)
  \right]
\mathrm{e}^{-\tau\mathcal{H}}.
\label{jx}
\end{equation}
Here  we also avoid analytical continuation, see, e.g., Refs.~\onlinecite{Trivedi96,Denteneer99,Mondaini12}.







\begin{figure}[t]
\includegraphics[scale=0.40]{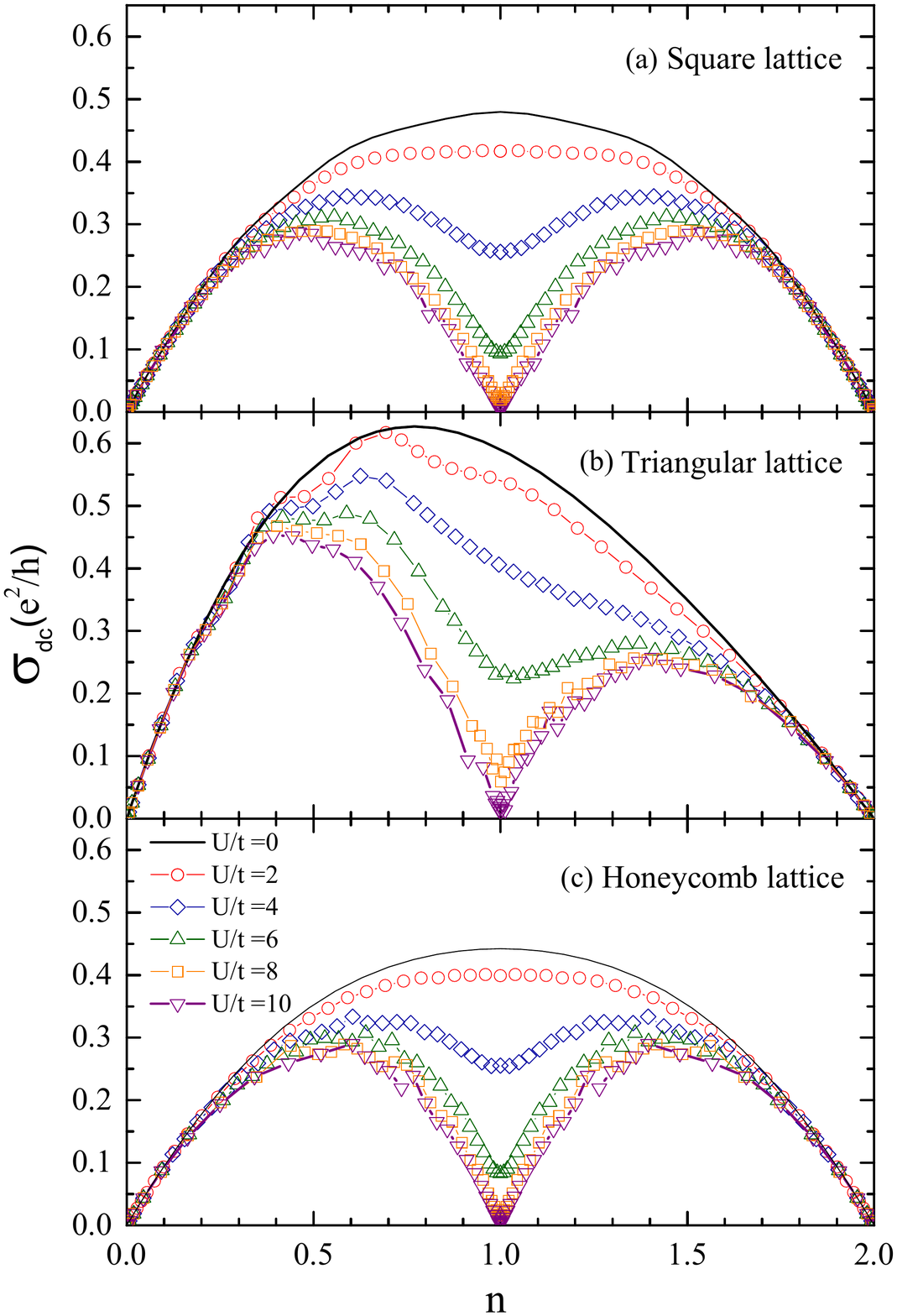}
\caption{(Color online) Longitudinal DC conductivity at $T/t=0.5$ as a function of density for the (a) square, (b) triangular and (c) honeycomb lattices. The dip at half-filling signifies the opening of the Mott gap, and the presence (absence) of bipartiteness is reflected in the particle-hole symmetric (asymmetric) behavior of the conductivity. The locations of the peaks of $N(\omega = 0)$ are very different from that of $\sigma_{DC}$ in the strongly interacting regime. Note that the dip in LDOS in Fig. \ref{fig:DOS} is not accompanied by a dip in $\sigma_{dc}$ in the honeycomb lattice.}
\label{fig:conductivity}
\end{figure}

Similar to what is seen for the LDOS, correlations reduce the conductivity, with the non-interacting conductivity as an upper limit for all the geometries studied as shown in Fig.\,\ref{fig:conductivity}.
Once again, correlations are shown to be irrelevant to transport properties for densities near the completely empty or filled bands, while its effects increase as half-filling is approached, with $\sigma_{dc} \to 0$ as $U/t$ increases.  It is interesting to note that, for the honeycomb lattice, the dip in the non-interacting DOS is not accompanied by a dip in the conductivity [shown in Figs.\,\ref{fig:DOS}(c) and \ref{fig:conductivity}(c), respectively].
For large values of $U/t$, the maxima for the conductivity are not at the same densities as the ones for the LDOS; for the square lattice, the conductivity has maxima at $n=0.5$ and $n=1.5$. For the honeycomb lattice, the maxima are at $n=0.6$ and $1.4$ and for the triangular lattice, the maxima are at $n=0.45$ and $n=1.45$.  For the triangular lattice, one can see that both the $U/t=0$ peak and the higher intensity peak for the LDOS, which are above half-filling, move to densities below half-filling for the conductivity.

\section{Seebeck coefficient}
\label{Sec_Seebeck}
\begin{figure}[t]
\includegraphics[scale=0.40]{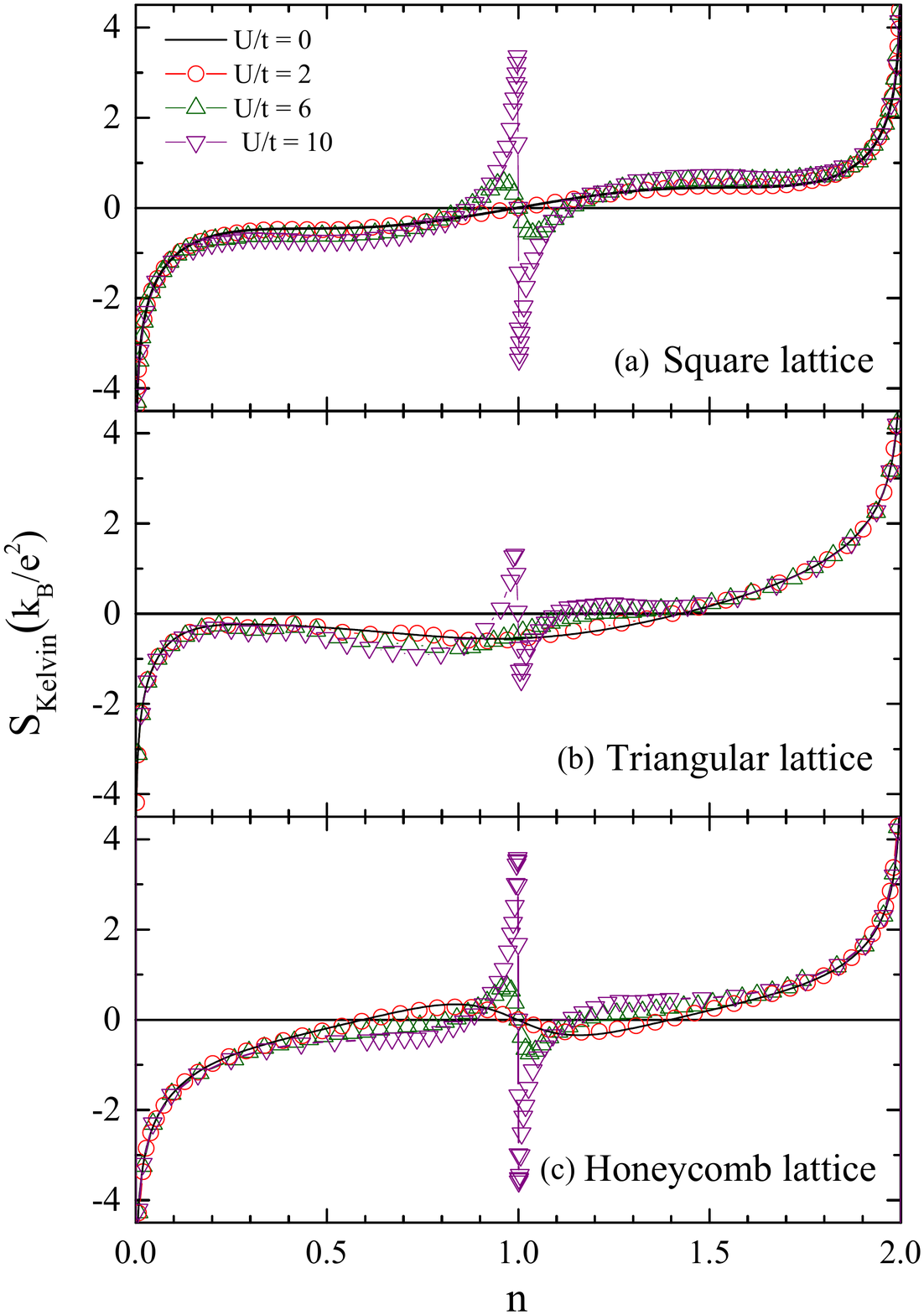}
\caption{(Color online) Seebeck coefficient for different values of the interaction strength at $T/t=0.5$ as a function of density for the (a) square, (b) triangular, and (c) honeycomb lattices. For clarity, we have reduced the set of $U/t$ values compared to previous plots. In the non-interacting limit, the Seebeck coefficient changes sign at half-filling for bipartite lattices (square and honeycomb), but at a finite doping for triangular lattice. With strong interactions, there is an enhancement of the Seebeck coefficient near half-filling, as well as an anomalous sign changes away from half-filling, signaling a change in carrier type. Note that for the honeycomb lattice, there is a sign change in Seebeck coefficient away from half filling even at $U/t = 0$, due to presence of Van Hove singularities from the Bloch bands.}
\label{fig:Seebeck}
\end{figure}

We now turn to the Seebeck coefficient, which is obtained from the entropy by using the Kelvin formula\,\cite{Peterson10,arsenault2013entropy},
\begin{align}
S_{\rm Kelvin} =  \left.  \frac{1}{e} {\frac{\partial s}{\partial n}}\right|_{T,V}~.
\label{Eq_kelvin_seebeck}
\end{align}
Within this approach, Fig.\,\ref{fig:Seebeck} displays the behavior of $S_{\rm Kelvin}$ in units of $k_B/e^2$ as a function of the electronic density for the (a) square, (b) triangular, and (c) honeycomb lattices.

At this point, it is worth recalling that the sign of the Seebeck coefficient is directly related to the type of carrier, being negative for holes and positive for electrons.  As a consequence of particle-hole symmetry for the square and honeycomb lattices, one has $S_{\rm Kelvin}(n)=-S_{\rm Kelvin}(2-n)$, leading to $S_{\rm Kelvin}(n=1)=0$. For all geometries examined, the effect of correlations is strongly dependent on density, being negligible for $n \lesssim 0.3$ and $n \gtrsim 1.7$, as previously observed for entropy, DOS and conductivity.

\begin{figure}[t]
\includegraphics[scale=0.34]{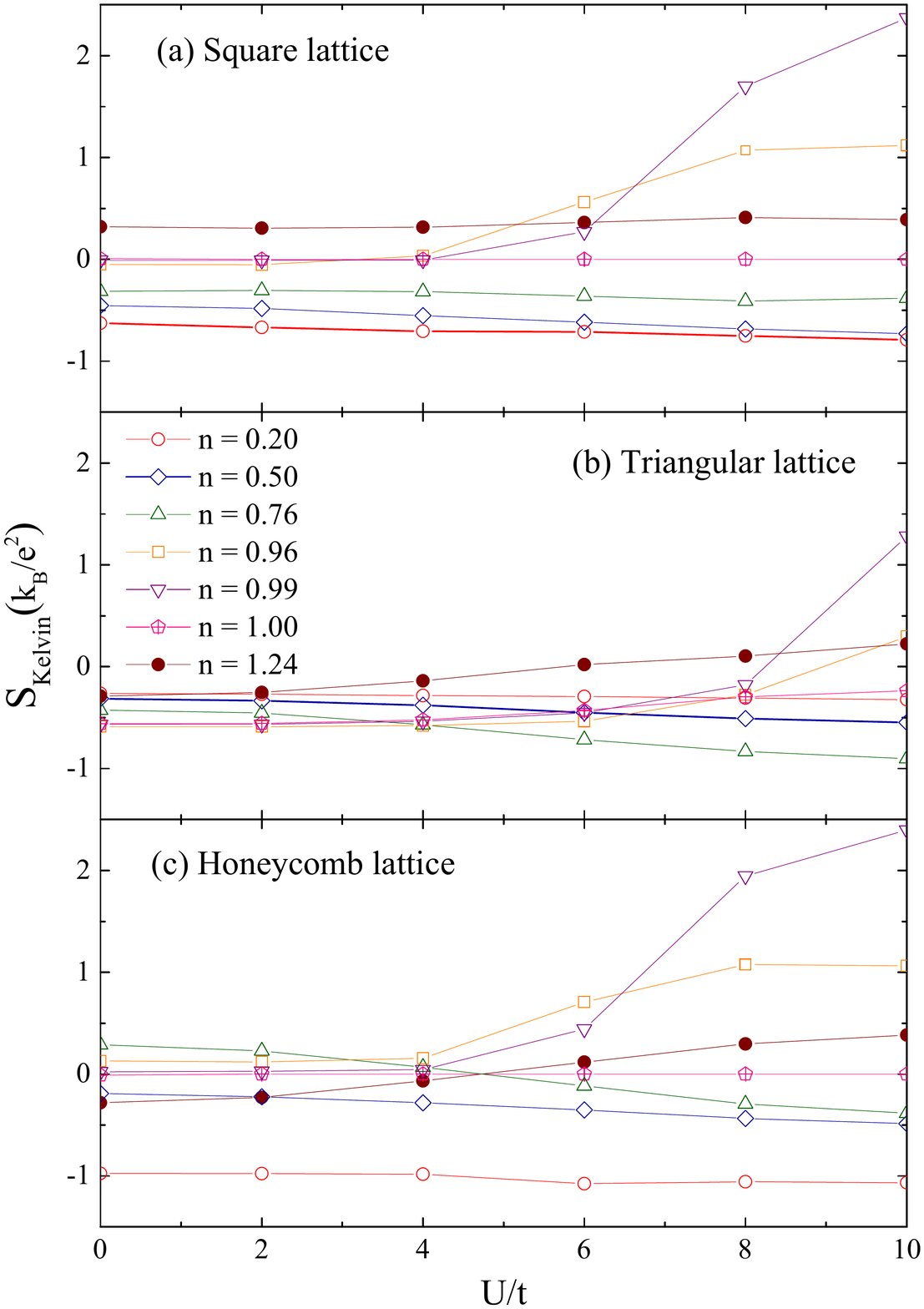}
\caption{(Color online) Seebeck coefficient as a function of $U/t$, at a fixed $T/t=0.5$ for different densities in (a) square, (b) triangular and (c) honeycomb lattices. For densities close to half-filling, the Seebeck coefficient changes sign and enhances as $U/t$ is increased.}
\label{fig:Seeback-U}
\end{figure}

 As expected, the Seebeck coefficient for the non-interacting square lattice presents only one sign change, at half-filling, as shown in Fig.\,\ref{fig:Seebeck}\,(a). 
 However, notice that in the presence of strong correlations ($U/t \gtrsim 6$) there is an anomalous behavior, characterized by a change of sign for densities away from half-filling, also displayed in Fig.\,\ref{fig:Seeback-U}\,(a) for $n \approx 0.96$. 
 In addition, there is a notable increase in the absolute value of Seebeck coefficient, compared to the non-interacting case. For instance, at $n=0.99$, $S_{\rm Kelvin} \approx 2.37 ~ k_B/e^2$, at $U/t = 10$, while for $U/t = 0$, $S_{\rm Kelvin} \approx -0.01 ~   k_B/e^2$. This steep increase can be explained by noting that the Seebeck coefficient, as defined in Eq.~\eqref{Eq:Seebeck}, is the ratio between the longitudinal thermoelectric and electrical conductivities. As we approach the Mott insulator at half-filling, electrons form local moments and electrical transport is strongly reduced, as shown by both conductivity and LDOS in the previous section. 
 The fast-decreasing electrical conductivity in the vicinity of half-filling must be accompanied by a non-vanishing  thermoelectric current for the peaks to form.
We can understand this as follows: at half-filling and strong correlations, each site is singly occupied and local moments are completely formed, there is no electric or thermoelectric transport and the Seebeck coefficient is zero. As we move slightly away from half-filling, there is a background of local moments over which $p=1-n$ carriers lead to thermoelectric and electric currents. This reduced number of carriers is in line with the breaking of Luttinger count \cite{Osborne,Hanke,Sakai} that has been established for the Fermi Hubbard model  and is in agreement with the change in carrier density in Hall experiments for YBCO\,\cite{badoux2016change}.
 
 
Due to absence of particle-hole symmetry in the triangular lattice, the sign change of $S_{\text{Kelvin}}$ in the non-interacting limit occurs away from half-filling,  at $n = 1.42$, as shown in Fig.\,\ref{fig:Seebeck}\,(b), as opposed to the square lattice, which has particle-hole symmetry, and exhibits the sign change symmetrically around half filling.
There is a range of densities, around  $0.5 \le n \le 0.9 $ where correlations lead to a small increase in modulus of the Seebeck coefficient.
In contrast to the square lattice, the peak in the Seebeck coefficient for the triangular lattice 
at $n=0.99$ is only present for very strong interactions. 
For $U/t= 8$, $S_{\rm Kelvin} \approx -0.18 ~ k_B/e^2 $, while for $U/t = 0$, for this electronic density  $S_{\rm Kelvin} \approx -0.55 ~ k_B/e^2 $. 
Strong correlations, around $U/t=10$, are needed to change the sign of the Seebeck coefficient at half-filling, as presented in Fig.\,\ref{fig:Seeback-U}\,(b).
Interestingly, above half-filling and below the sign-change point for $U/t=0$, the correlations are detrimental to the thermopower up to $U/t=8$, bringing the Seebeck coefficient closer to zero.

Fig.\,\ref{fig:Seebeck}\,(c) shows the Seebeck coefficient for the honeycomb lattice.
For the non-interacting system, $S_{\rm Kelvin}$ changes sign at  $n = 0.6$, 1.0, and 1.4, therefore the charge of the carriers is negative for $n < 0.6$ and $1.0 < n < 1.4$, while it is positive for $0.6 < n < 1.0$ and $n> 1.4$.
Interestingly, these densities correspond to the peaks and dip positions for the conductivity, as seen in Fig.\,\ref{fig:conductivity}\,(c).
We recall that such changes in carriers come from the fact that the honeycomb lattice has two bands for the noninteracting case, so one may expect changes from electron to hole properties of the transport coefficients depending on the filling of each band. 
However, correlations push the sign change to values closer to half-filling, going from $n=0.6$ and $1.4$ at $U/t=0$ to $n=0.85$ and $1.15$ at $U/t=8$. One thing to note is that although the behavior of the Seebeck coefficient is very different between the square and the honeycomb lattice in the noninteracting limit, adding interactions and transitioning to the Mott insulating state seems to wash these differences away. This can be understood by noting that in the noninteracting picture, the transport is determined by the Bloch bands which are different for square and honeycomb lattices. However, with strong interactions, Mott physics destroys the Bloch bands, opens up a Mott gap, and forms upper and lower Hubbard bands. While details of the single particle bands are erased, the information about the particle-hole symmetry imprinted in the many body spectrum is still retained. 
Hence the behavior of the Seebeck coefficient is qualitatively similar between the square and honeycomb lattice but different from the triangular lattice.

Similar to the square lattice, there is a significant increase in the Seebeck coefficient for the honeycomb lattice close to half-filling at $U/t = 6$, 8 and 10, as shown for $n=0.96$ and $n=0.99$ in Fig.\,\ref{fig:Seeback-U}\,(c). For $U/t = 10$, for example, we have that for $n = 0.99$, $S_ {\rm Kelvin} \approx 2.30 ~ k_B/e^2$, while in the non-interacting case ($U/t = 0$) for this density value, $S_{\rm Kelvin} \approx 0.03 k_B/e^2$, an enhancement of two orders of magnitude.



\begin{figure}[t]
\includegraphics[scale=0.40]{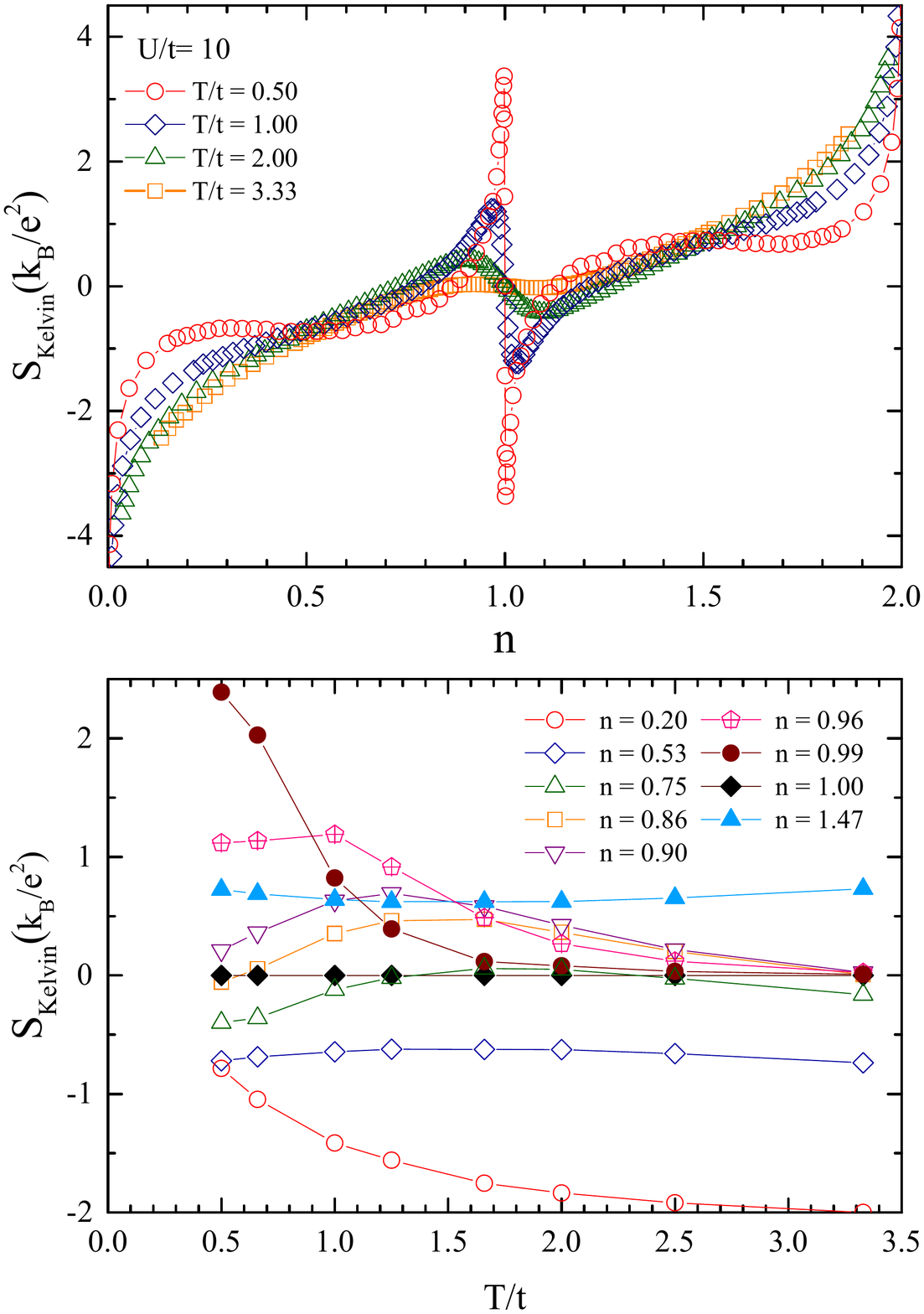}
\caption{(Color online) Seebeck coefficient for the square lattice at $U/t=10$ as (a) a function of density for the square lattice at different temperatures (b) a function of temperature for different densities. In panel (a), the Seebeck coefficient shows anomalous behavior at low temperatures, and approaches the free particle limit as $T/t$ is increased. In panel (b), at low densities ($n=0.2$), the Seebeck coefficient has the expected sign, and monotonically decreases with temperature. At a critical density $n \approx 0.5$ (also shown for $n \approx 1.5$ in the electron doped side), the Seebeck coefficient is almost temperature independent. 
In the anomalous region ($0.75 \lessapprox n \lessapprox 1.0$), peaks in the Seebeck coefficient start to move towards the smallest temperature considered here, as one moves towards the half-filling limit, where it becomes 0 for all temperatures.   }
\label{fig:Seeback-U10}
\end{figure}

We also analyze the effects of temperature on the thermopower for the square lattice at $U/t=10$.
Figure \ref{fig:Seeback-U10}\,(a) displays $S_{\rm Kelvin}$ as a function of $n$ for different $T/t$, while Fig.\,\ref{fig:Seeback-U10}\,(b) shows $S_{\rm Kelvin}$ as a function of $T/t$ for different densities.
Curves for different temperatures nearly cross around $n=0.5$ and $n=1.5$ [Fig.\,\ref{fig:Seeback-U10}(a)], leading to almost horizontal lines for $n=0.53$ and $n=1.47$ in Fig.\,\ref{fig:Seeback-U10}(b).
For $n \lesssim 0.5$ and $n \gtrsim 1.5$ reducing the temperature is detrimental to the thermopower, as seen by a reduction of the modulus of the Seebeck coefficient for $n=0.2$.
For densities in the range $0.5 \lesssim n \lesssim 0.9$ and $1.1 \lesssim n \lesssim 1.5$, the behavior of $S_{\rm Kelvin}$ with temperature is non-monotonic, and can change sign with $T/t$,  as shown for $n=0.75$ in figure \ref{fig:Seeback-U10}(b).
Finally, the anomalous behavior in the vicinity of half-filling has a marked dependence on temperature, increasing the modulus of the Seebeck coefficient as $T/t$ is reduced. This fast increase of the thermopower with decreasing temperature close to half-filling has also been observed in the $t-J$ model\cite{peterson2010kelvin}, Hubbard model on an FCC lattice\cite{arsenault2013entropy} and $t-t^\prime-U$ Hubbard Model \cite{Deveraux}.

\section{Thermoelectric Power factor}
\label{Sec_Power_factor}
Proceeding, we now discuss the effects of correlations to the thermoelectric Power Factor (\textit{PF}), defined as 
\begin{equation}
    PF= S^2 \sigma,
\end{equation}
where the dc conductivity ($\sigma$) and the Seebeck coefficient (\textit{S}) were obtained in Sections~\ref{DOS_conductivity} and \,\ref{Sec_Seebeck}, respectively. Simultaneously increasing both the modulus of the Seebeck coefficient and the conductivity maximizes the Power Factor. Strategies to determine the Seebeck coefficient that leads to an optimum thermoelectric Power Factor have been sought theoretically for systems that can be described by the Boltzmann transport equation\,\cite{PF-1} and experimentally for  CZTS thin films\,\cite{PF-2} and La-doped SrTiO$_3$\,\cite{PF-3}
 thin films.

\begin{figure}[t]
\includegraphics[scale=0.35]{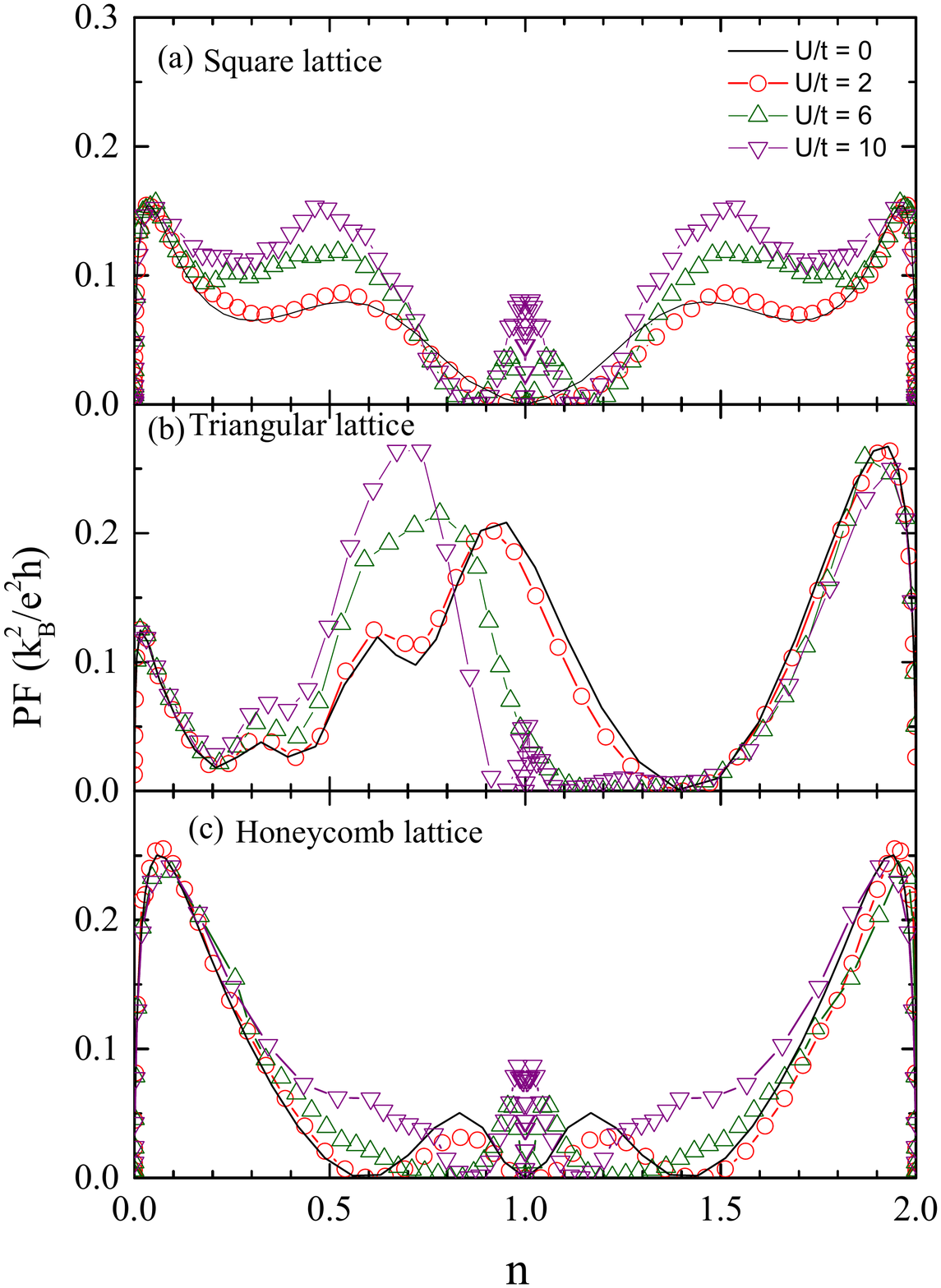}
\caption{(Color online) Thermoelectric Power factor as a function of density for (a) square, (b) triangular and (c) honeycomb lattices. In the free particle limit (almost empty and almost filled lattice), the Power factor exhibits a peak for all values of $U/t$. Interaction causes additional peaks to develop very close to the Mott insulating limit for large interaction strengths. Additional features for intermediate doping also appear that are strongly influenced by lattice geometry.}
\label{fig:PF}
\end{figure}

Fig.~\ref{fig:PF} shows the thermoelectric Power Factor as a function of density for fixed $T/t=0.5$, and for the three analyzed geometries. A common feature to all data is the presence of peaks close to the empty ($n \lesssim 0.1$) and completely filled lattices ($n \gtrsim 1.9$). Starting with completely empty (filled) lattices, as density increases (decreases) the conductivity increases, and the Seebeck coefficient modulus decreases, leading to peaks that are independent of interaction strength. 
As the conductivity goes to zero for Mott insulators, $PF \to 0$ at half-filling for a geometry-dependent value of $U/t$.  For the square lattice, the hump in $PF$ for $U/t=0$ around $n  \approx 0.6$ (1.4) turns into a peak at $n \approx 0.5$ (1.5) with the increase of correlations, the dominant contribution coming from the conductivity.

The effect of correlations for the intermediate densities peak of the honeycomb lattice is more subtle, as can be seen in figure \ref{fig:PF}(c). The peaks for the non-interacting system at $n=0.85$  ($n=1.15$) decrease in intensity with correlations for $U/t=2$ and 4, and then a shoulder develops at lower (higher) densities for larger $U/t$. Close to half-filling once again correlations drive the Seebeck coefficient peak up which in turn leads to the $PF$ peaks.

For the triangular lattice one can see that the non-interacting system has a peak at $n=0.95$. Correlations move the peak to lower densities and increase its intensity, as clearly seen in figure \ref{fig:PF}(b), where the peak for $U/t=10$ is at $n \simeq 0.7$. As for the other geometries, a peak develops in the vicinity of the Mott insulating state, here only seen for the larger values of $U/t$ studied.

Comparing the overall effects of correlations in the different geometries analyzed, we see that there are clear correlation-induced peaks near half-filling in all cases. For the square and honeycomb lattices, the non-interacting thermoelectric power factor is zero at half-filling and is driven to  $PF  \approx 0.1 \  k_B^2/e^2 h$. Around quarter and three-quarter fillings as well, correlations also play a relevant role in increasing the power factor for the square and honeycomb lattices. Correlations are very effective in increasing the Power Factor in a triangular lattice, where  $PF  \simeq 0.27 \ k_B^2/e^2 h$, the largest value obtained, for $U/t=10$ at $n \sim 0.7$. For this geometry, we observe that density can be used to tune the thermoelectric Power Factor.


\section{Conclusions}
\label{conclu}
We have studied the thermoelectric properties of strongly interacting two dimensional systems with different geometries. Our results clearly show an anomaly in the Seebeck coefficient in the vicinity of half-filling, characterized by an enhanced response depending on both geometry and interaction strength. The anomaly is characterized by a change in the sign of the carriers which is accompanied by an interaction-induced increase. The anomaly is also intensified with the reduction of temperature. 

The thermoelectric Power Factor displays a competition between the Seebeck coefficient and the conductivity. The anomaly in the Seebeck coefficient is reflected in the $PF$, with correlation-driven peaks immediately below and above half-filling at geometry-dependent values of $U/t$.  The decreasing conductivity near half-filling is the limiting factor in the intensity of  $PF$ in this region of densities. Away from half-filling,  at intermediate densities (around $n=0.4-0.6$ and $n=1.4-1.6$) the peaks in $PF$ have a strong contribution from the conductivity with positions  strongly dependent on geometry. For this range of densities, peak position and intensity can be tuned by correlations. Although the Seebeck coefficient is smaller for the triangular lattice, the Power Factor for this geometry shows the higher peak values and the stronger tunability with density and correlations, making it a strong candidate for enhanced thermoelectric properties. 

\section*{ACKNOWLEDGMENTS}

Financial support from 
Fundação Carlos Chagas Filho de Amparo à Pesquisa do Estado do Rio de
Janeiro, grant numbers E-26/204.308/2021 (W.C.F.S.),  E-26/200.258/2023 - SEI-260003/000623/2023 (N.C.C.),  E-26/200.959/2022 (T.P.), and E-26/210.100/2023 (T. P.); CNPq grant numbers  313065/2021-7 (N.C.C.), 403130/2021-2 (T.P.), and 308335/2019-8 (T.P.) is gratefully acknowledged.
 We also acknowledge support from INCT-IQ. We acknowledge support from NSF Materials Research Science and Engineering Center (MRSEC) Grant No. DMR-2011876 and NSF-DMR 2138905 (SR,AS,NT). We also acknowledge computational resources from the Unity cluster at the Ohio State University.

\color{black}

\bibliography{references.bib}

\end{document}